\documentclass[aip,apl,reprint]{revtex4-1}

\usepackage{graphicx}
\usepackage{dcolumn}
\usepackage{bm}
\usepackage{url}
\usepackage{sistyle}
\usepackage{float}
\usepackage{mathtools}
\usepackage{color}

\begin{document}

\title{Characterization of Hydrogen Plasma Defined Graphene Edges} 

\author{Mirko~K.~Rehmann}
\thanks{M.~K.~R. and Y.~B.~K. contributed equally to this work}
\affiliation{Department of Physics, University of Basel, CH-4056 Basel, Switzerland}

\author{Yemliha~B.~Kalyoncu}
\thanks{M.~K.~R. and Y.~B.~K. contributed equally to this work}
\affiliation{Department of Physics, University of Basel, CH-4056 Basel, Switzerland}

\author{Marcin~Kisiel}
\affiliation{Department of Physics, University of Basel, CH-4056 Basel, Switzerland}

\author{Nikola~Pascher}
\affiliation{Nanosurf AG, Gr\"aubernstrasse 12, 4410 Liestal, Switzerland}

\author{Franz~J.~Giessibl}
\affiliation{Department of Physics, University of Regensburg, 93053 Regensburg, Germany}

\author{Fabian~M\"uller}
\affiliation{Department of Physics, University of Basel, CH-4056 Basel, Switzerland}

\author{Kenji~Watanabe}
\affiliation{National Institute for Material Science, 1-1 Namiki, Tsukuba 305-0044, Japan}

\author{Takashi~Taniguchi}
\affiliation{National Institute for Material Science, 1-1 Namiki, Tsukuba 305-0044, Japan}

\author{Ernst~Meyer}
\affiliation{Department of Physics, University of Basel, CH-4056 Basel, Switzerland}

\author{Ming-Hao~Liu}
\email[]{minghao.liu@phys.ncku.edu.tw}
\affiliation{Department of Physics, National Cheng Kung University, Tainan 70101, Taiwan}

\author{Dominik~M.~Zumb\"uhl}
\email[]{dominik.zumbuhl@unibas.ch}
\affiliation{Department of Physics, University of Basel, CH-4056 Basel, Switzerland}

\date{\today}

\begin{abstract}
We investigate the quality of hydrogen plasma defined graphene edges by Raman spectroscopy, atomic resolution AFM and low temperature electronic transport measurements. The exposure of graphite samples to a remote hydrogen plasma leads to the formation of hexagonal shaped etch pits, reflecting the anisotropy of the etch. Atomic resolution AFM reveals that the sides of these hexagons are oriented along the zigzag direction of the graphite crystal lattice and the absence of the D-peak in the Raman spectrum indicates that the edges are high quality zigzag edges. In a second step of the experiment, we investigate hexagon edges created in single layer graphene on hexagonal boron nitride and find a substantial D-peak intensity. Polarization dependent Raman measurements reveal that hydrogen plasma defined edges consist of a mixture of zigzag and armchair segments. Furthermore, electronic transport measurements were performed on hydrogen plasma defined graphene nanoribbons which indicate a high quality of the bulk but a relatively low edge quality, in agreement with the Raman data. These findings are supported by tight-binding transport simulations. Hence, further optimization of the hydrogen plasma etching technique is required to obtain pure crystalline graphene edges. 

\end{abstract}

\maketitle

Graphene edges play an important role in many physical phenomena\cite{fujita1996peculiar,nakada1996edge}. In particular, the edge termination has a strong influence on the electronic properties of graphene nanoribbons (GNRs). Crystallographic edges of the armchair (AC) type are predicted to enable the creation of helical modes and Majorana fermions\cite{klinovaja2013giant} and were proposed as candidates for the implementation of spin qubits\cite{trauzettel2007spin}. For pure zigzag (ZZ) edges, edge-magnetism was predicted to emerge which can be used for spin filtering\cite{son2006half}. For these effects to be observable in experiment, high quality edges are necessary because edge disorder suppresses magnetic correlations\cite{yazyev2010emergence} and leads to electron localization which complicates transport studies\cite{mucciolo2009conductance,oostinga2010magnetotransport,stampfer2009energy,gallagher2010disorder,liu2009electrostatic,molitor2010energy}. It has been observed, that GNRs fabricated by e-beam lithography and reactive ion etching (RIE) in an Ar/O$_2$ plasma have a high degree of edge disorder\cite{oostinga2010magnetotransport,stampfer2009energy,gallagher2010disorder,liu2009electrostatic,molitor2010energy}. Hence, other approaches to create GNRs with high quality edges are pursued such as carbon nanotube unzipping\cite{jiao2009narrow,jiao2010facile,kosynkin2009longitudinal}, ultrasonication of intercalated graphite\cite{li2008chemically}, chemical bottom up synthesis\cite{ruffieux2016surface,cai2010atomically}, anisotropic etching by nickel nanoparticles\cite{campos2009anisotropic}, anisotropic etching during CVD processing\cite{geng2013fractal,guo2015governing,zhang2011anisotropic,stehle2017anisotropic}, or carbothermal etching of graphene sheets\cite{nemes2010crystallographically,krauss2010raman,oberhuber2013weak,oberhuber2017anisotropic,oberhuber2017anisotropic}. Another promising approach which was considered to create high quality crystallographic graphene edges is to employ a hydrogen (H) plasma to perform anisotropic etching of graphite and graphene\cite{mccarroll1971reactivity,mccarroll1970interaction,yang2010anisotropic,shi2011patterning,xie2010selective,wang2016patterning,diankov2013extreme,wu2018magnetotransport,zhang2012experimentally,hug2017anisotropic}.

In this study, we characterize H plasma defined graphene edges on graphite and single layer (SL) graphene on hexagonal boron nitride (hBN) by means of atomic force microscopy (AFM), Raman spectroscopy and low temperature electronic transport measurements. We find high quality ZZ edges on graphite surfaces, manifested by the absence of the D-peak in the Raman spectrum\cite{cancado2004influence,basko2009boundary}. In contrast, SL graphene on hBN edges exhibit a large D-peak which is indicative of the presence of edge disorder and AC segments. In comparison, the D-peak intensity measured at H plasma defined edges is twice as large as on edges created with RIE. Polarization dependent Raman measurements reveal an edge configuration which consists of approx. $60\,$\% ZZ and $40\,$\% AC segments. Moreover, electronic transport measurements performed across a \textit{pnp} junction of a H plasma treated graphene flake exhibit Fabry-P\'{e}rot oscillations, reflecting the high electronic quality of the bulk graphene flake after H plasma exposure. However, at high magnetic field valley-isospin oscillations appear and indicate a rather low edge quality. In a second device we investigate transport through narrow GNRs with RIE defined edges and H plasma defined edges and find comparable mobilities for these two edge types.

The results from the Raman experiments and the electronic transport studies give a consistent picture, indicating the presence of disorder at H plasma defined graphene edges and thus the need for optimization of the etching process to enable the creation of high quality ZZ edges.

\subsection{Results and Discussion}

\begin{figure}[ht]
\includegraphics[width=8.6cm]{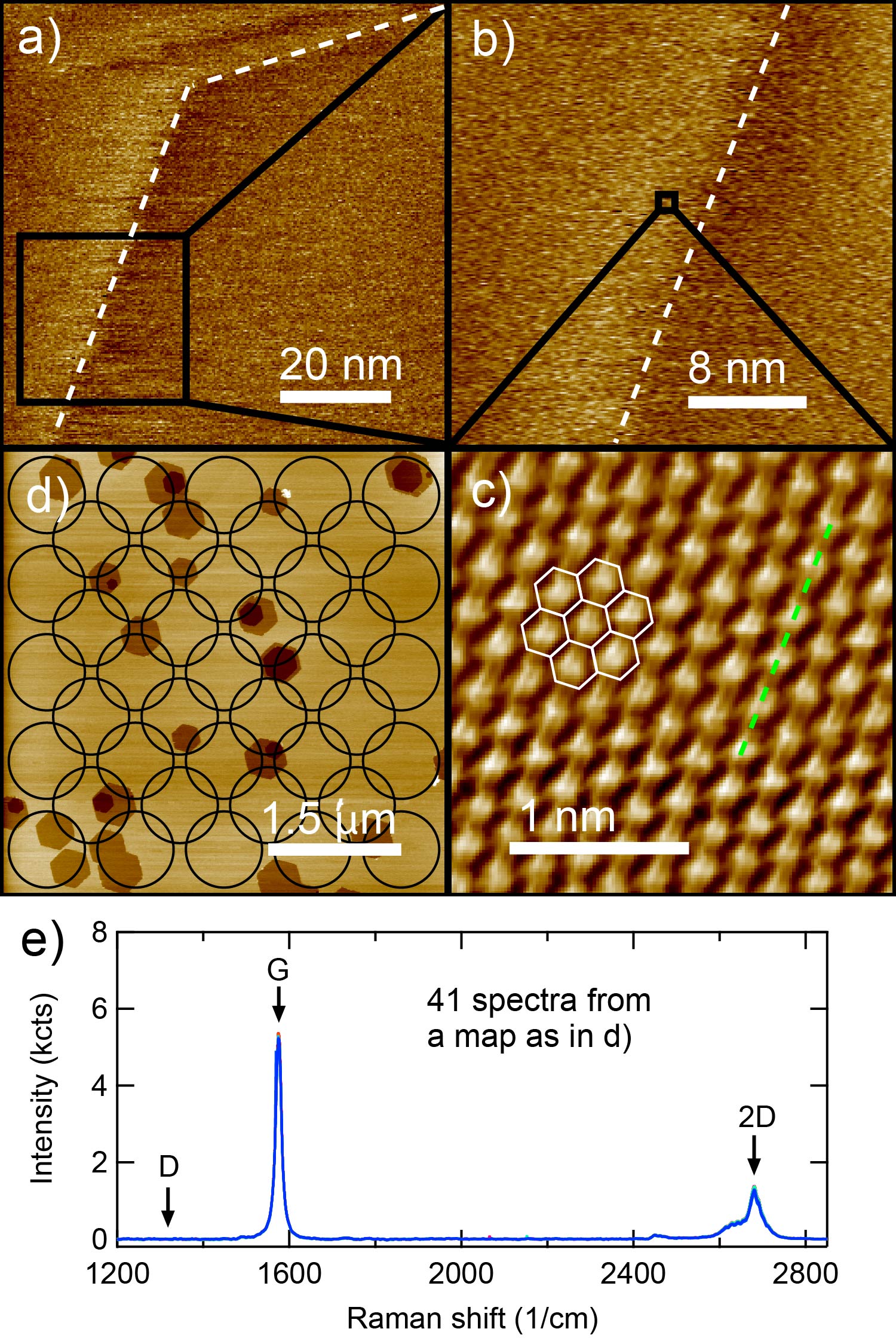}
\vspace{-5mm}
\caption{\label{fig:figure1}{\bf AFM maps and Raman spectra of H plasma etched graphite} (a) AFM height image of a section of a hexagonal shaped etch pit on a graphite flake which was exposed to a remote hydrogen plasma. (b) Zoom-in on data shown in panel a. (c) Atomic resolution constant height AFM force image of the black squared region in b. The graphene lattice is superimposed in white. The green dashed line indicates the ZZ direction and is parallel to the hexagon edges (white dashed lines in a and b). (d) Tapping mode AFM image of a $5\times5\,\mathrm{\mu m^2}$ area of a graphite flake. The black circles with a diameter of $800\,$nm, given by the laser spot size, indicate the locations at which Raman spectra were taken. (e) 41 Raman spectra laid on top of each other, all recorded with circularly polarized light.}
\vspace{-1mm}
\end{figure}

\textbf{High Quality ZZ Edges on Graphite.} In a first step of the experiment, we intend to visualize the edge of a hexagon created by H plasma exposure, to learn its crystallographic direction and its atomic configuration. Therefore, we record topography and force images by means of ambient qPlus based atomically resolving AFM\cite{wastl2013optimizing} (Figure\,\ref{fig:figure1}a - c) on a graphite surface which shows several hexagons. In Figure\,\ref{fig:figure1}a we show one corner of a hexagon, and its edges are demarcated with white dashed lines. From a to c, black squared regions are scanned with higher resolution with the same sample orientation. Figure\,\ref{fig:figure1}c is a constant height atomic resolution force image of the graphite surface close to the edge and the hexagonal lattice structure is superimposed on the image. The green dashed line in Figure\,\ref{fig:figure1}c is drawn parallel to the white dashed lines in Figure\,\ref{fig:figure1}a and b. This picture clearly shows that the edge is parallel to the ZZ direction, in agreement with recent findings\cite{yang2010anisotropic}. Although the above discussed AFM measurements allow to unambiguously assign the macroscopic edge orientation of the hexagons to the ZZ direction, thermal drift hindered to position the edge inside the scan range for the atomic resolution imaging, hence it was not possible to actually visualize the atomic configuration of the hexagon edge.

To access information about the edge configuration on the atomic level, Raman measurements are conducted on a graphite flake which was exposed to H plasma under similar conditions as the sample investigated in Figure\,\ref{fig:figure1}a to c. 41 Raman spectra are taken over a $5\times5\,\mathrm{\mu m^2}$ region, covering the whole area as shown in Figure\,\ref{fig:figure1}d with the grid of black circles. This web of spectra makes sure that the surface is fully covered. The resulting 41 spectra are laid on top of each other in Figure\,\ref{fig:figure1}e. These 41 spectra are essentially the same; in particular, the G and 2D-peaks fit to each other and there is no D-peak as seen in Figure\,1e. The absence of the D-peak in any of these spectra indicates high quality ZZ edges since any edge disorder would result in some D-peak intensity\cite{cancado2004influence,basko2009boundary,gupta2008probing,you2008edge,casiraghi2009raman}.

\textbf{Raman Spectroscopy on SL Graphene Hexagons on hBN.} Next, we investigate the edge quality of hexagons created in SL graphene flakes on a hBN substrate. In a previous work\cite{hug2017anisotropic}, we showed that the character of the etch is substrate dependent and that it is also possible to get highly anisotropic etching if SL graphene is placed on a hBN substrate. However, it remains unclear how good the edge quality is on a microscopic level. To find out, we prepare a SL graphene flake on a hBN substrate and etch it for $4\,$h in a remote H plasma to perform Raman measurements at the created graphene edges.

\begin{figure}[t]
\includegraphics[width=8.6cm]{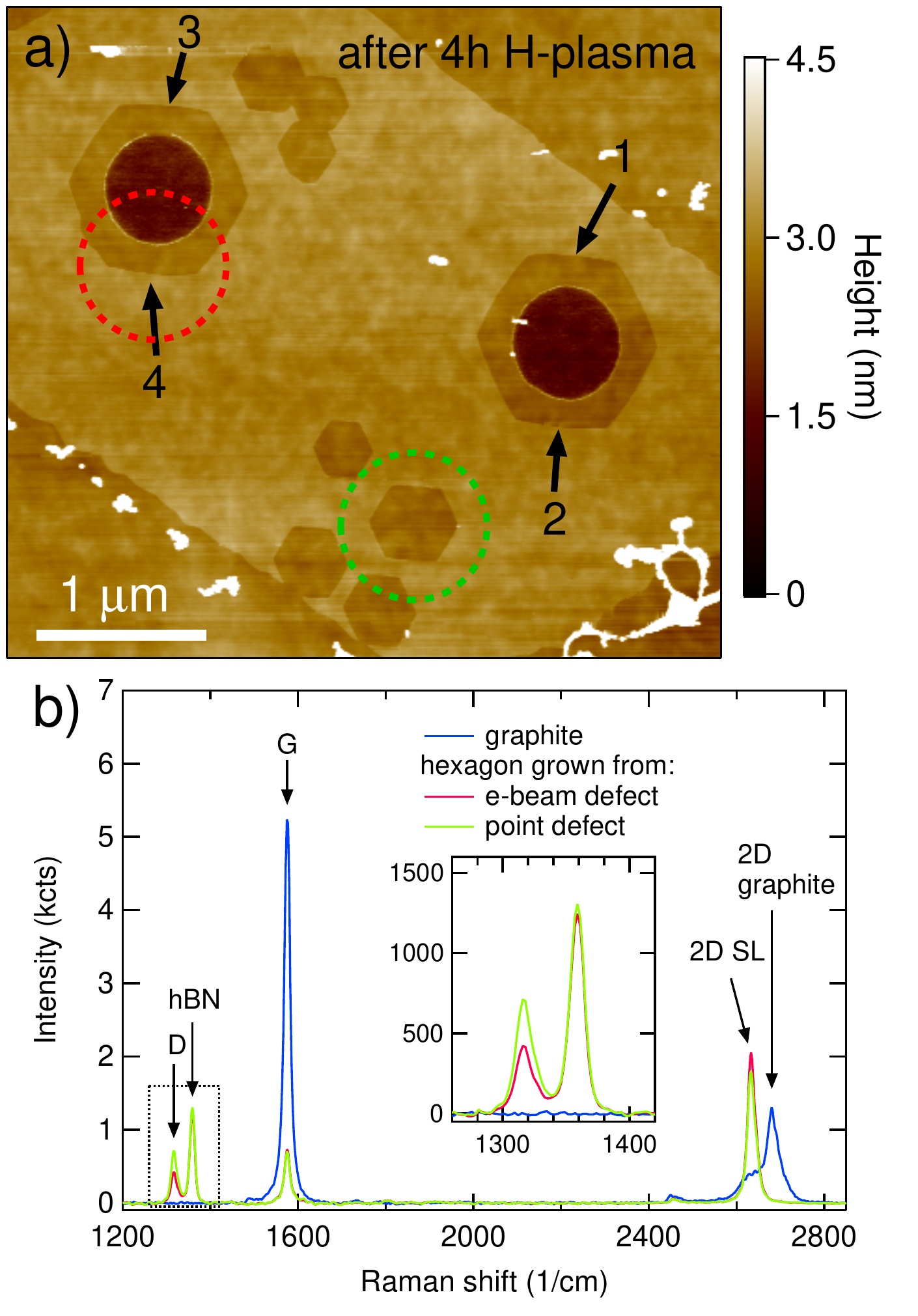}
\vspace{-5mm}
\caption{\label{fig:figure2} {\bf AFM height image and Raman spectra of H plasma defined SL graphene edges} (a) AFM height image of a SL graphene flake on a hBN substrate after $4\,$h of remote H plasma exposure. Two round shaped defects of a diameter of $600\,$nm were created by e-beam lithography and RIE etching in a Ar/O$_2$ plasma. They serve as nucleation centers for the anisotropic etch which transforms them into hexagonal etch pits. Besides the two patterned defects, there are defects which grow into the smaller hexagons visible next to the larger ones. The red and green dashed circles indicate the locations at which the Raman spectra shown in b were recorded. The black numbers denote the different investigated edge segments of which the measurements are shown in Figure\,\ref{fig:figure3}f. (b) Raman spectra of graphite (blue) and of SL graphene edges encircled by the green and red dashed circles in panel a. The inset shows the region of the D-peak. All spectra are recorded with circularly polarized light.}
\vspace{-1mm}
\end{figure}

Figure\,\ref{fig:figure2}a shows an AFM height image of a SL graphene flake on a hBN substrate. The two darker disks are induced defects which we fabricated by means of e-beam lithography and RIE in an Ar/O$_2$ plasma. Upon H plasma exposure, they transform into regular hexagonal shaped etch pits. Moreover, smaller hexagons grow next to the two big ones. Those smaller hexagons are either grown from lattice defects already present after exfoliation or induced during H plasma exposure (i.e. by highly energetic ions). To learn about the edge quality of such SL graphene edges, we record Raman spectra at the locations indicated by the red and green dashed circles in Figure\,\ref{fig:figure2}a and show them in Figure\,\ref{fig:figure2}b. We observe the graphene related G and 2D-peaks and the hBN peak coming from the substrate. More importantly, the graphene D-peak appears to the left of the hBN peak at both measurement locations. The difference in intensity could stem from differences of the probed edge segment length. Further, we overlay a spectrum taken on the graphite sample shown in Figure\,\ref{fig:figure1} (blue curve). Apparently, there is no D-peak for the graphite case whereas we do observe a D-peak for the SL graphene on hBN edges. This indicates, that there is a significant amount of disorder present at the hexagon edges in SL graphene.

\indent Obviously, the hexagon edges created in SL graphene on hBN are of different quality compared to the edges of hexagons formed on graphite. Already in our previous study\cite{hug2017anisotropic} we have observed that the substrate has a big influence on the etching character. Although hBN as a substrate allows for highly anisotropic etching, the edge configuration on a microscopic level is different from the one on graphite surfaces. This could be due to several reasons, e.g. the different lattice constants of graphene and hBN could potentially lead to strain effects\cite{couto2014random} or to the appearance of Moir\'{e} superlattice effects\cite{xue2011scanning} which could influence the quality of the H plasma etching process.

\begin{figure*}[ht]
\includegraphics[width=18cm]{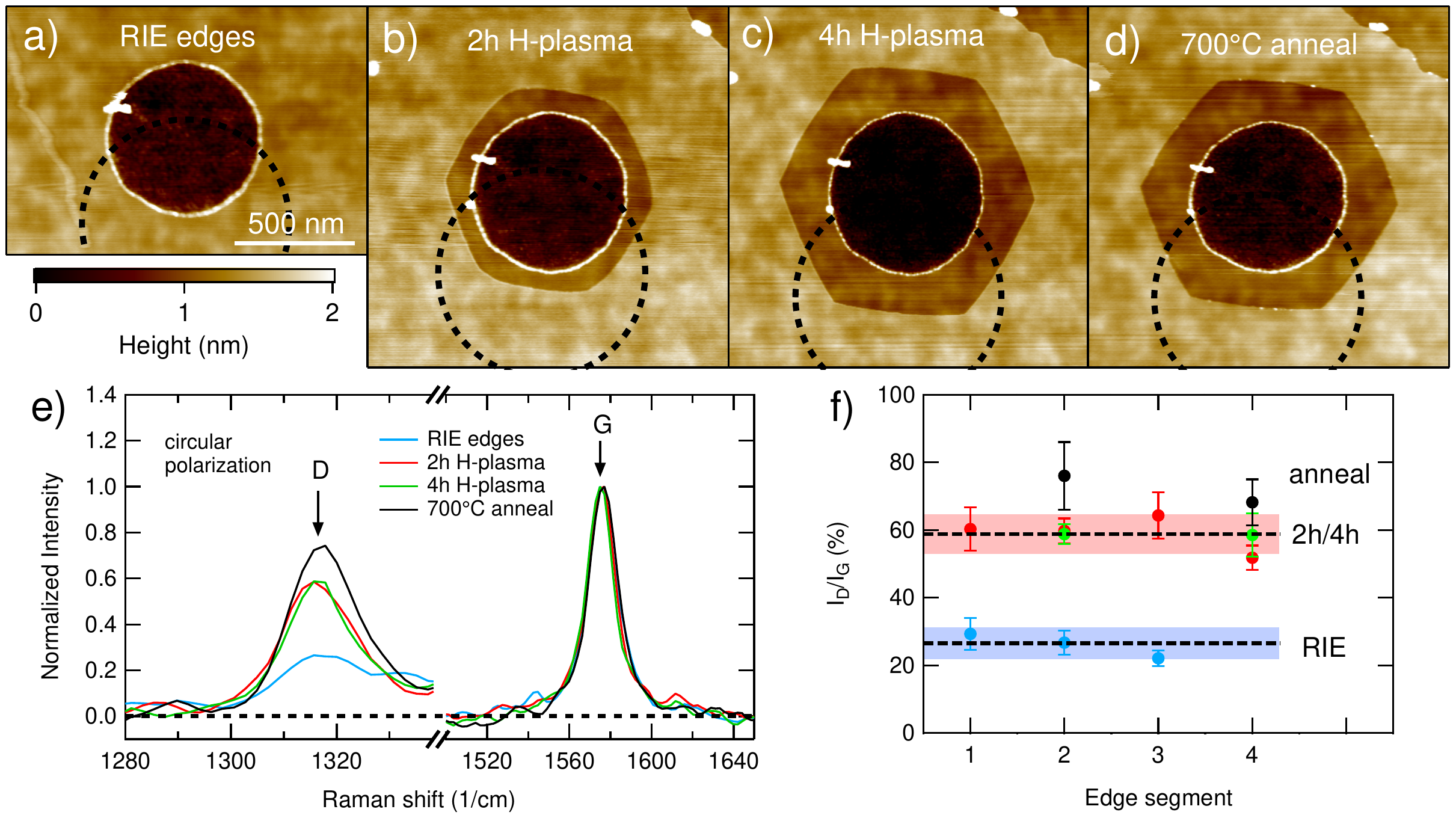}
\vspace{-7mm}\caption{\label{fig:figure3}{\bf Comparison of RIE defined edges with H plasma defined edges} (a) to (d) AFM height images of e-beam defined defects in SL graphene on hBN, after RIE a, after $2\,$h b and $4\,$h c of remote H plasma etching and after annealing at $T=700\,\degC$ for 30$\,min$ at a pressure of $1.6\cdot 10^{-3}\,$mbar d. The black dashed circles indicate the spot size of the Raman measurements. (e) Raman spectra recorded with circularly polarized light at the bottom edge of the right hole (edge segment \#2, see Figure\,\ref{fig:figure2}a after RIE, $2\,$h, $4\,$h and after annealing at $T=700\,\degC$. The spectra are normalized to the G-peak and each curve is an average of five measurements. (f) Normalized D-peak intensities recorded at different edge segments as labeled in Figure\,\ref{fig:figure2}a. The blue and red shaded bands are the standard deviations from all the corresponding measurements.}
\vspace{-1mm}
\end{figure*}

\textbf{Evolution of the Raman D-mode from RIE to H Plasma Defined Graphene Edges.} Next, we study the evolution of the observed D-peak over a time sequence of the etching process. This series of measurements shows how the edge quality evolves from a RIE defined circular hole to H plasma defined edges and further studies the effect of annealing. We started with defining circular holes by means of e-beam lithography and RIE with an Ar/O$_2$ plasma, which creates disordered edges without any defined crystal orientation\cite{oostinga2010magnetotransport,stampfer2009energy,gallagher2010disorder,liu2009electrostatic,molitor2010energy,bischoff2011raman}. The AFM image of this RIE defined circular hole is shown in Figure\,\ref{fig:figure3}a. Raman single spectra were recorded with circularly polarized light at the bottom edge of the hole indicated by the black dashed circle. Circular polarization ensures that the Raman signal is collected equally at every point of the edge, regardless of the edge direction. After the measurements, the sample is exposed to the remote H plasma first for 2 hours and then 2 more hours, creating hexagonal etch pits of increasing diameter as shown in Figure\,\ref{fig:figure3}b and c. As a final step, we annealed the sample in vacuum. After each step, Raman spectra are measured at the same location.

In Figure\,\ref{fig:figure3}e, four Raman spectra measured at different stages shown in a to d are plotted, normalized to the G-peak height. All spectra are averages over five measurements recorded under same conditions and looking all very similar. As expected, the RIE defined hole shows a D-peak (blue curve). After the first H plasma etching (red curve) the D-peak intensity surprisingly increases approximately by a factor of two and stays at this level for further etching (green curve). Finally, annealing at an elevated temperature of 700\,$\degC$ again increases the D-peak (black curve), suggesting structural edge defects as the D-peak origin, since annealing likely would reduce, not increase, the amount of edge hydrogenation. We note also that when investigating bulk graphene with Raman spectroscopy, where no edge segments are inside the laser spot, we do not observe any D-peak, consistent with previous work \cite{elias2009control}, indicating the absence of bulk hydrogenation (see supplementary online material (SOM) S1), though it is in principle possible that annealing at even higher temperatures might be required to remove hydrogen from the edge. The bright circular rim of the RIE defined circular hole on hBN (Figure\,\ref{fig:figure3}b - d) does not contribute to the graphene bands in the Raman spectrum (see SOM S1). Further, changes in both graphene area and edge length enclosed in the laser spot are giving negligible contributions to the evolution of the D-peak; see SOM S5 for details. In Figure\,\ref{fig:figure3}f we find the same trend in D-peak intensity for all four different edge segments which are indicated in Figure\,\ref{fig:figure2}a by the black arrows and numbers. The values for different edge segments stay in a narrow window, giving consistent results.

The increase of the D-peak upon the first H plasma exposure could stem from the formation of AC segments at the edges, since AC edges are highly D-peak active\cite{cancado2004influence,basko2009boundary,gupta2008probing,you2008edge,casiraghi2009raman}. From previous studies\cite{hug2017anisotropic}, it is clear that the direction of the edge generally goes along the ZZ direction. Hence, we conclude that the SL graphene edges on hBN run along the ZZ direction but have a substantial amount of disorder, probably at least partially in form of AC segments. To test this hypothesis, we study the edge disorder with the angular dependence using linearly polarized light\cite{casiraghi2009raman,ThermalDynamics}.

\begin{figure}[ht]
\includegraphics[width=8.6cm]{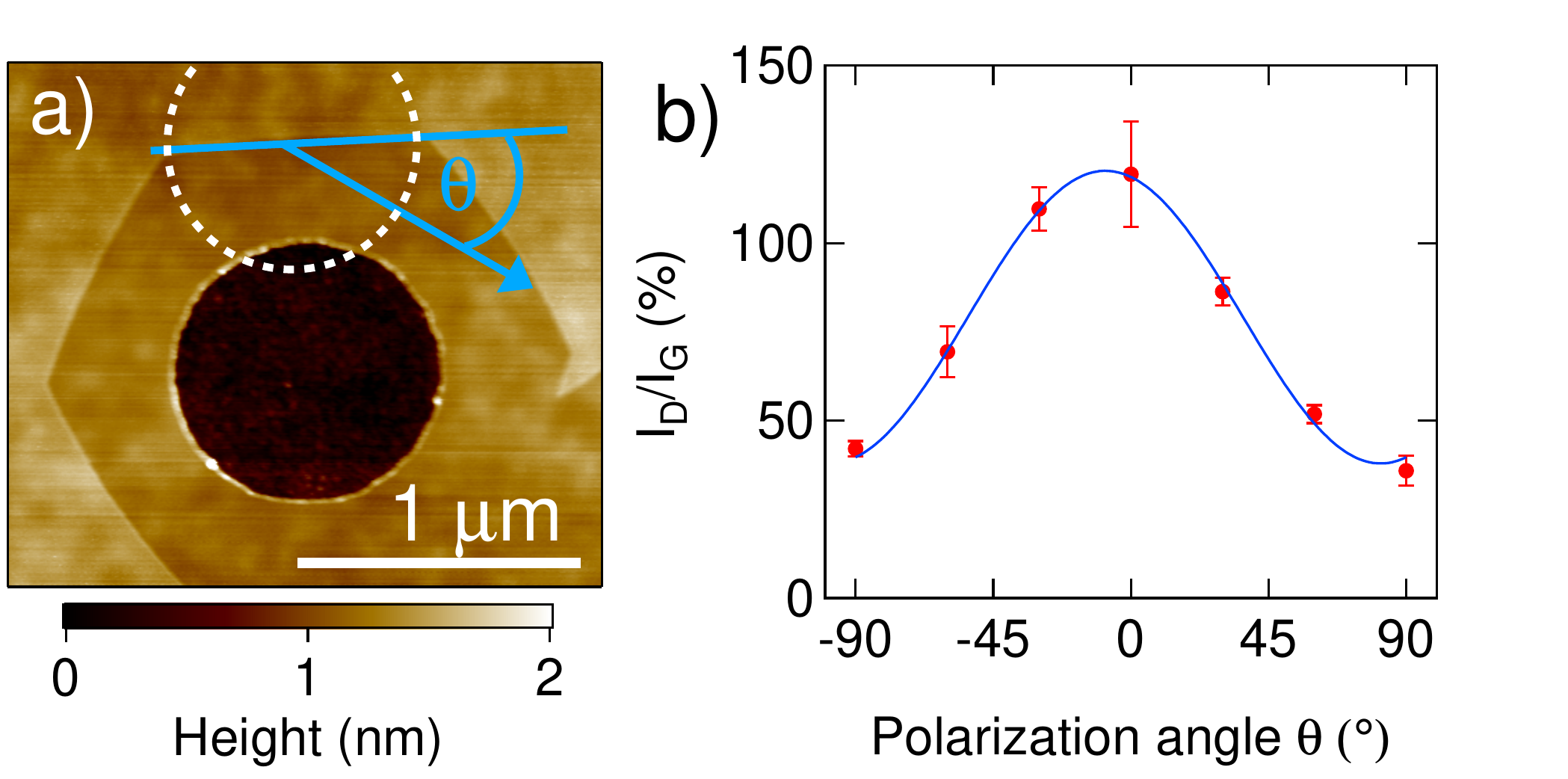}
\vspace{-5mm}\caption{\label{fig:figure4} {\bf Polarization angle dependence of a SL/hBN graphene edge} (a) AFM height image of a hexagonal etch pit in a SL graphene flake on a hBN substrate after $6\,$h of remote H plasma exposure. The white dashed circle indicates the laser spot where the Raman spectra were recorded and $\theta$ denotes the angle of the laser light polarization with respect to the graphene edge. (b) Normalized D-peak intensity for different polarization angles $\theta$. The blue curve is a fit to equation\,1 of ref.\,\onlinecite{casiraghi2009raman} yielding $I(D)_{min}=38\pm2$, $I(D)_{max}=120\pm2$ and $\theta_{max}=-8\pm1$.}
\vspace{-1mm}
\end{figure}

\textbf{Polarization Angle Dependent Raman Measurements.} Xu and coworkers in ref. \onlinecite{ThermalDynamics} have observed edge reconstruction on ZZ edges due to thermal treatment. Since we etch our samples at a temperature of $T=400\,\degC$, it might be that also our graphene edges experience thermal reconstruction. Indeed, it is theoretically predicted that an AC edge has lower energy compared to a ZZ edge\cite{koskinen2008theory}. A model to extract the relative abundance of AC-30$^\circ$ segments and point defects was proposed in ref. \onlinecite{ThermalDynamics}. We apply this model to our data and see that the observed D-peak signal only comes from AC-30$^\circ$ segments and that essentially no point defects are present (see SOM S4). Furthermore, Casiraghi et al. in ref. \onlinecite{casiraghi2009raman} proposed a theory to calculate the ratio of ZZ segments to AC-30$^\circ$ segments which we apply to our data; see Figure\,\ref{fig:figure4}. Figure\,\ref{fig:figure4}a shows an AFM height image of a hexagon created in SL graphene supported on a hBN substrate. The Raman spectra were recorded at the laser spot indicated by the white dashed circle. The angle $\theta$ of the polarization with respect to the edge is marked in light blue. In Figure\,\ref{fig:figure4}b we plot the normalized D-peak intensities as a function of $\theta$. The blue curve is a fit to equation\,1 of ref.\,\onlinecite{casiraghi2009raman}. Since our hexagons exhibit rather straight edges, we take an equal amount of +30$^\circ$ and -30$^\circ$ AC segments. We find that our graphene edge consists of about $59\pm2\,\%$ ZZ and $41\pm2\,\%$ AC-30$^\circ$ segments. This is in excellent agreement with a second data set acquired on a different hexagon on the same graphene flake (see SOM S4 for data taken at different stages of the etching process).

Besides the polarization dependence of the D-peak, the G-peak can also serve to get insight into the structure of graphene edges\cite{sasaki2010identifying,cong2010raman}. In particular, a clean AC edge is expected to exhibit a cos$^2(\theta)$ dependence and a clean ZZ edge a sin$^2(\theta)$ dependence. An edge with a mixture of ZZ and AC segments would result in a corresponding mixed angular dependence with a weak amplitude of modulation. This is what is seen in our data and is thus again consistent with a similar mixture of ZZ and AC segments.\\

\begin{figure*}[ht]
\includegraphics[width=18cm]{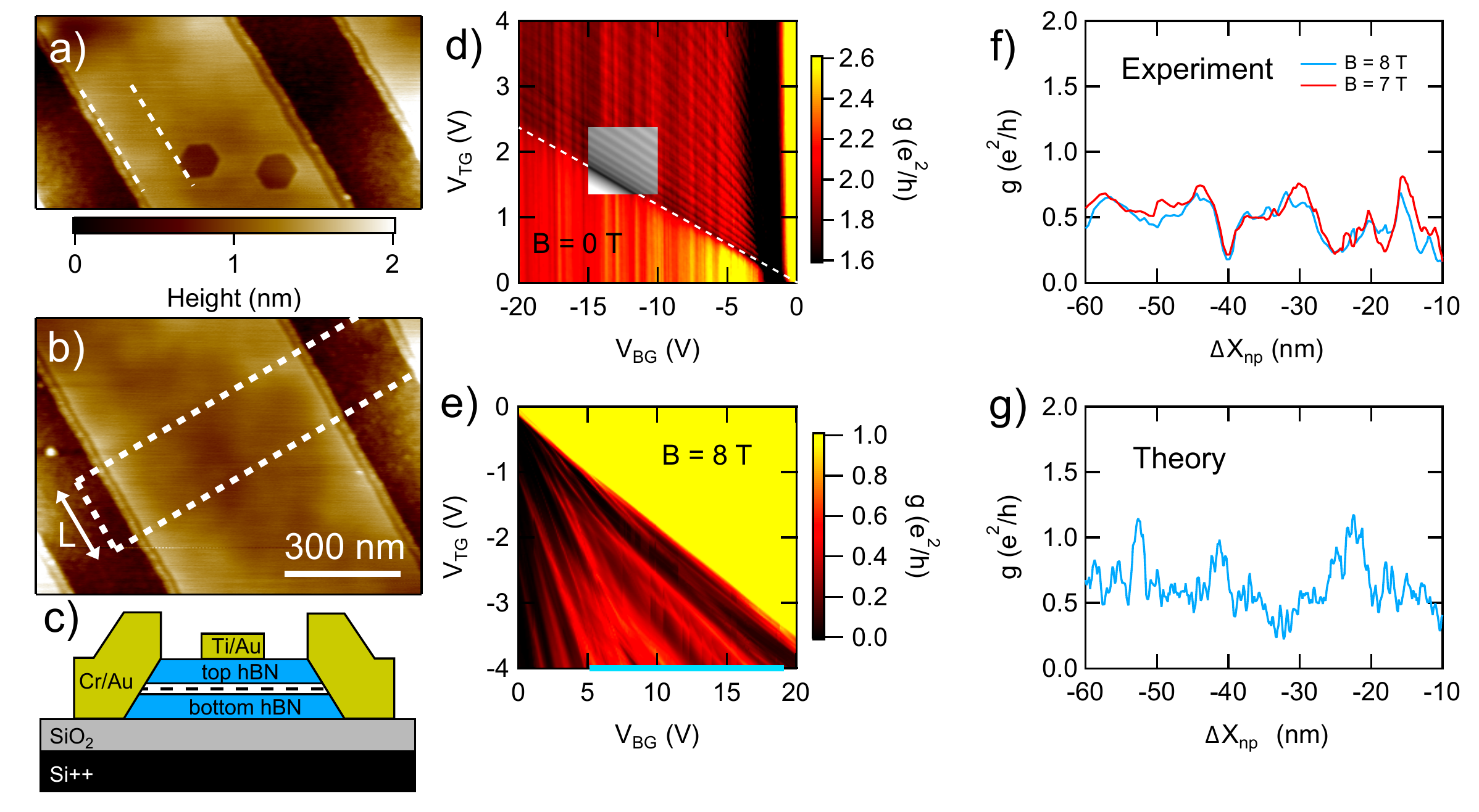}
\vspace{-7mm}\caption{\label{fig:figure5}{\bf Electronic transport measurements of encapsulated GNR with H plasma defined edges} (a) AFM height image of a SL GNR. The white dashed lines indicate that the hexagon edges are well aligned with the GNR edges. (b) AFM height image of the GNR on which electronic transport was measured. The white dashed lines indicate the location of the $200\,$nm top gate which was evaporated on top of a hBN capping layer. The length $L$ of the charge carrier cavity is tunable with gate voltages. (c) Device schematic of the encapsulated GNR with a global back gate and a local top gate. The black dashed line indicates the SL GNR. (d) Differential conductance as a function of back gate V$_\text{BG}$ and top gate V$_\text{TG}$ voltage at $B=0\,$T in the \textit{pnp} region (\textit{n} under the top gate). The greyscale inset is obtained by a simulation and matches very well with the experiment. The white dashed line marks the charge neutrality point in the dual gated regime. (e) Similar map as in d but recorded at $B=8\,$T and in the \textit{npn} regime. In the bi-polar regime, resonances fanning out linearly from the charge neutrality point are visible. (f) Cut along the blue line in e and an additional curve recorded at same gate voltages but at $B=7\,$T. The x-axis was converted from V$_\text{BG}$ to the \textit{np}-interface location relative to the physical top gate edge $\Delta X_\text{np}$; see SOM S7 and S8 for details. (g) Calculation of the conductance through a GNR following the ZZ direction with disorder in form of AC-30$^\circ$ segments and a bulk disorder of $35\,$meV plotted versus $\Delta X_\text{np}$.}
\vspace{-1mm}
\end{figure*}

\indent \textbf{Fabry-P\'{e}rot Interference in a H Plasma Defined GNR \textit{pnp} Junction.} Next, we investigate the influence of H plasma treatment on the electronic property of graphene. In particular, we address the quality of bulk graphene and features arising from the H plasma defined graphene edges. To this end, we fabricate a SL GNR with H plasma defined edges following the ZZ direction of the crystal lattice, done as follows. After a first exposure of the graphene flake to the remote H plasma, a few hexagons grow from which we can learn the crystallographic orientation of the flake. Next, the graphene flake is cut into stripes which run parallel to the hexagon sides and hence parallel to the ZZ direction of the crystal lattice (see Figure\,\ref{fig:figure5}a). Subsequently, another H plasma exposure leads to etching from the ribbon edges and thus leaves a GNR with H plasma defined graphene edges. We note that the investigated GNR is free of defects in form of missing carbon atoms, because otherwise these defects would have grown into hexagons (see Figure\,\ref{fig:figure5}b). The white dashed rectangle in Figure\,\ref{fig:figure5}b indicates the location of the top gate which was fabricated after encapsulation with hBN\cite{wang2013one}. In Figure\,\ref{fig:figure5}c, a schematic of the cross section of the device is shown.

The local top gate and the global back gate allow to tune the charge carrier densities inside and outside the top gated regions individually. This enables the possibility to tune the system into bipolar regimes, i.e. \textit{pnp} or \textit{npn}, thus creating two \textit{pn} junctions which can form a resonance cavity for the charge carriers. If the charge carriers move ballistically inside such a cavity, Fabry-P\'{e}rot resonances appear\cite{campos2012quantum,young2009quantum,rickhaus2013ballistic}. In Figure\,\ref{fig:figure5}d, we show the conductance as a function of back gate voltage V$_\text{BG}$ and top gate voltage V$_\text{TG}$ in the \textit{pnp} regime, at zero magnetic field. Clear fringes due to Fabry-P\'{e}rot resonances are seen to be parallel to the diagonal white dashed line which marks the zero carrier density in the dual-gated region. This indicates that the observed fringes come from Fabry-P\'{e}rot interferences within the top-gated cavity. The grayscale inset overlaid on Figure\,\ref{fig:figure5}d is obtained by a quantum transport simulation based on an infinitely wide graphene lattice \cite{rickhaus2013ballistic,liu2012efficient}, with the electrostatically simulated barrier profile following the geometry of the device implemented. The simulation is obtained without fit parameters and matches very well with the experiment.

The oscillation frequency of the Fabry-P\'{e}rot resonances is linked to the cavity length $L$. We extract values for $L$ in the range of $L=160\,$nm to $330\,$nm (see SOM S3 for details about the cavity length extraction). Since $L$ represents a lower bound for the mean free path, we can calculate the corresponding lower bound of the mobility, which is approximately $60\,000\,$cm$^2$/Vs, reflecting the high electronic quality of graphene after H plasma exposure.

\textbf{Valley-Isospin Dependent Conductance Oscillations in a H Plasma Defined GNR.} We continue with high-field magnetotransport measurements, and follow an analysis based on ref. \onlinecite{isospin} to study the edge quality of our GNR sample. At high magnetic field, the edge state of the lowest Landau level is valley-isospin-polarized depending on the edge of the GNR\cite{tworzydlo2007valley}. By tuning the gate voltages, the zero-density region, i.e., the \textit{pn} and \textit{np} interfaces, can be controllably moved along the edges of the GNR, revealing the edge-specific conductance oscillations\cite{isospin}.

The conductance map $g(V_\text{BG},V_\text{TG})$ at $B=8\,$T is shown in Figure\,\ref{fig:figure5}e, where the fringes fanning out from the origin seen in the \textit{npn} regime (bottom left part) behave qualitatively similar as those reported in ref. \onlinecite{isospin}, indicating the emergence of the valley-isospin physics. Similar behavior is also found in the \textit{pnp} regime (not shown). In Figure\,\ref{fig:figure5}f, we show a cut (blue curve) from the map of Figure\,\ref{fig:figure5}e at $V_\text{TG} = -4\,$V (marked by the blue line) showing weak oscillations around $\sim 0.5\,$e$^2/$h. We re-interpret the back gate voltage values in terms of $\Delta x_\text{np}$: the change of the position of the gate-defined \textit{np} interface with respect to the left edge of the top gate based on our electrostatic simulation (see SOM S7 and S8). Another cut (red curve) recorded at $B=7\,$T shows a rather similarly oscillating curve, indicating that the oscillations have already developed at lower field. This further gives a sign that the observed oscillations originate from the edge-specific valley-isospin.

\begin{figure}[ht]
\includegraphics[width=8.6cm]{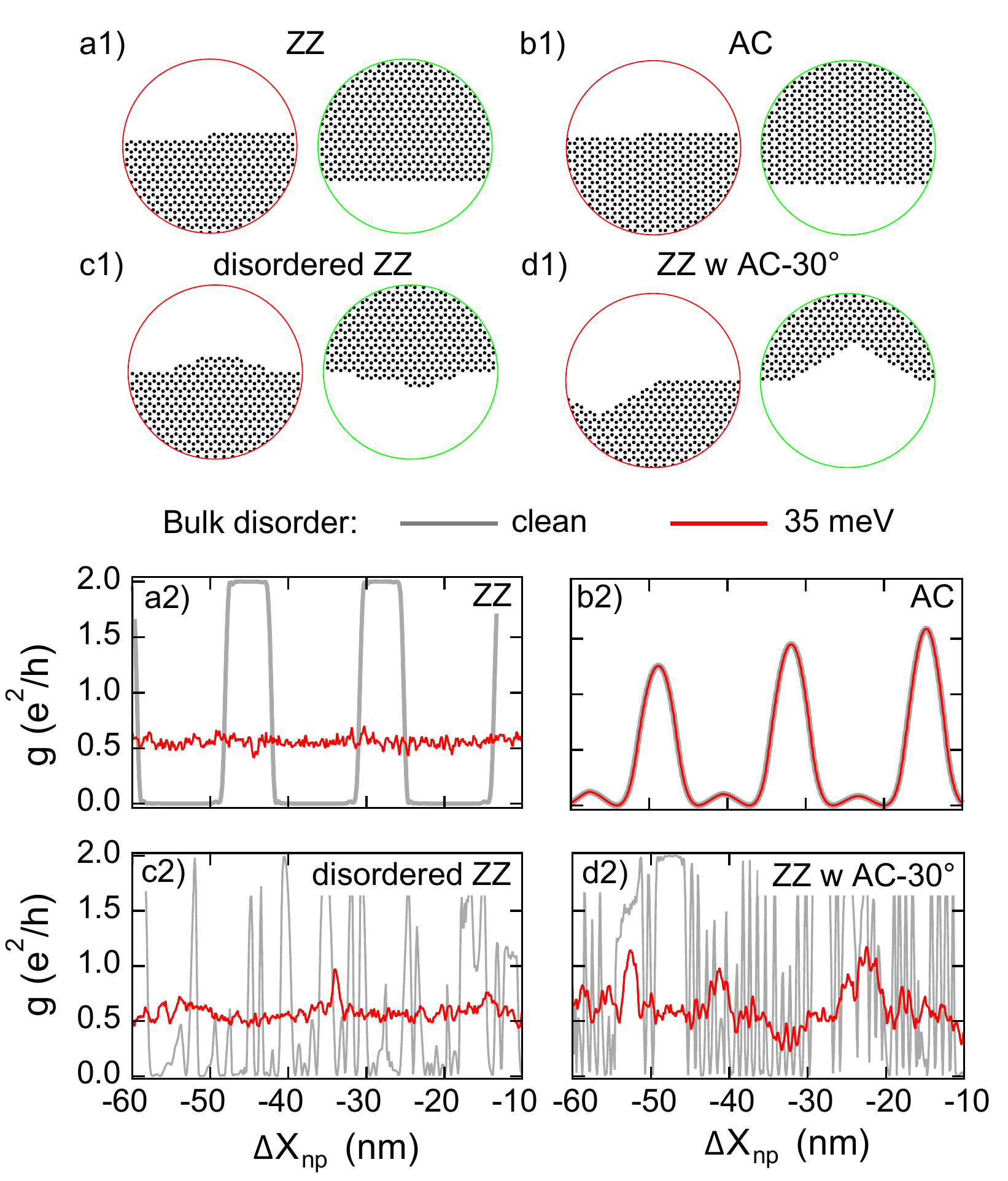}
\vspace{-5mm}\caption{\label{fig:figure6}{\bf Quantum transport simulations} (a1) Ribbon with a perfect ZZ edge at the bottom and one-atom steps at the top edge. (b1) Similar case as in a1 but following the AC direction. (c1) Ribbon following the ZZ direction but with a large amount of edge disorder on both edges. (d1) ZZ edge ribbon with 40\% fraction of randomly distributed AC-30$^\circ$ segments on both sides (as found in experiment) with a depth of $1\,$nm. The red and green circles are zoom-ins on the top and bottom edges of the ribbons,respectively. (a2) to (d2) Conductance as a function of $\Delta X_\text{np}$ with and without disorder, as labeled. Only the ZZ with AC-30$^\circ$ segments case qualitatively agrees with the experiment.}
\vspace{-1mm}
\end{figure}

To further confirm the origin of the experimentally observed conductance oscillations shown in Figure\,\ref{fig:figure5}f, we have performed quantum transport simulations for both AC and ZZ edges with various types of disorder, see Figure\,\ref{fig:figure6}; see SOM S8 for details.

Ideal ZZ or AC GNRs with perfect edges result in a constant conductance. In presence of edge disorder with one-atom steps on either only one edge (see Figure\,\ref{fig:figure6}a1 and b1) or on both edges (see Figure\,\ref{fig:figure6}c1 and d1), the conductance alternates between 0 and $\sim2\,$e$^2$/h whenever the np-interface sweeps across a single atom step for both ZZ (Figure\,\ref{fig:figure6}a2) and AC (Figure\,\ref{fig:figure6}b2) edges with zero bulk disorder. Stronger edge disorder with more frequent steps at the edges results in a correspondingly higher frequency of oscillation of the conductance, see Figure\,\ref{fig:figure6}c2 and d2. When a realistic {\em bulk} disorder of $35\,$meV strength is added (extracted from the experiment based on the width of the Dirac peaks), the oscillations in the ZZ case collapse to a roughly constant conductance of $\sim 0.5\,$e$^2$/h, while the conductance in the AC case remains strongly oscillating. Both of these behaviors are not consistent with the experiment. A ribbon which follows the ZZ direction with disorder in form of AC-30$^\circ$ segments shows oscillations in the simulations which are relatively robust against bulk disorder and which look qualitatively similar to the ones found in experiment (see Figure\,\ref{fig:figure5}g and Figure\,\ref{fig:figure6}d2). Hence, these findings are in agreement with AFM data indicating edges along the ZZ direction and are also in agreement with the Raman results presented above, i.e. graphene edges on hBN created by remote H plasma exposure follow the ZZ direction but contain significant edge disorder in form of AC-30$^\circ$ segments.

\begin{figure}[t]
\includegraphics[width=8.6cm]{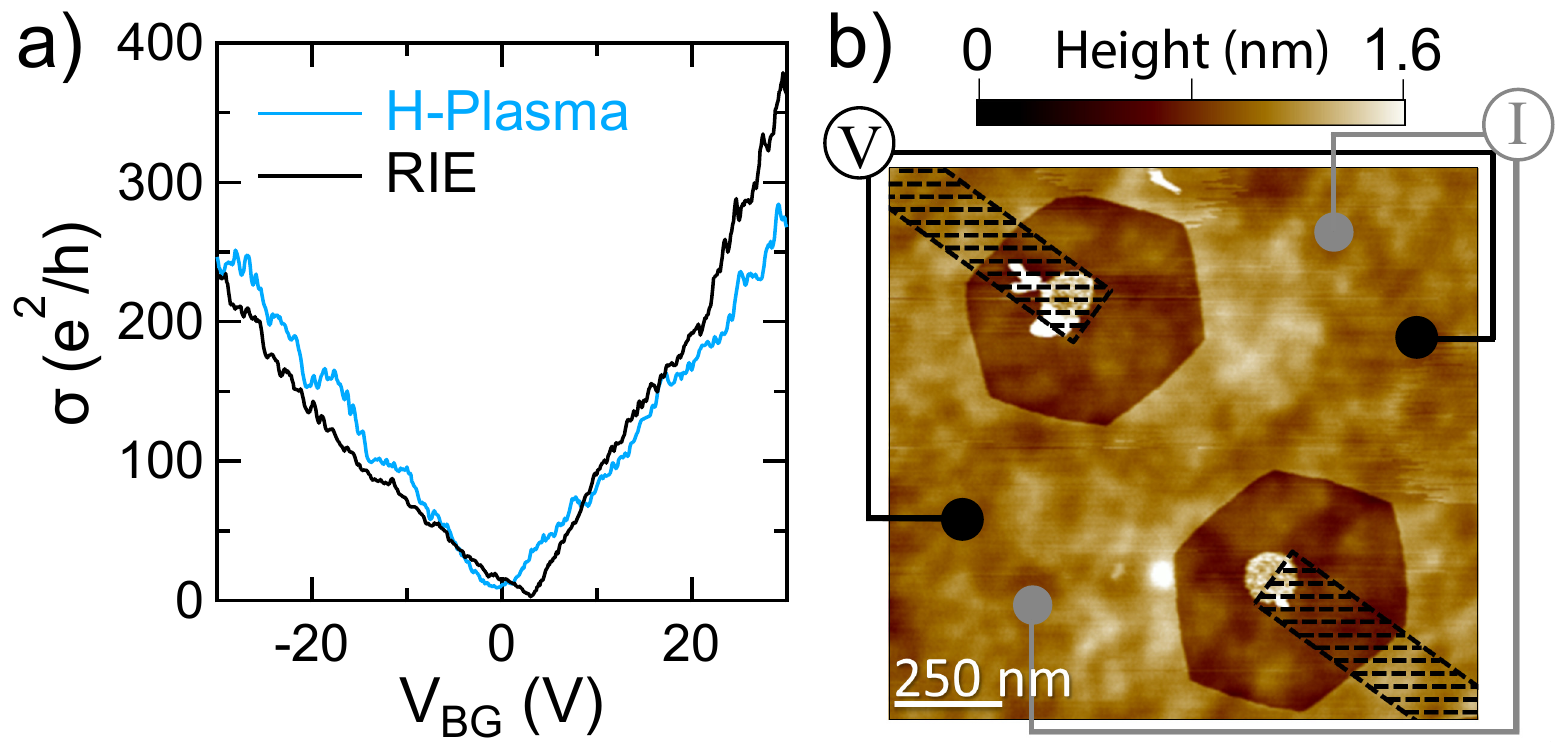}
\vspace{-5mm}\caption{\label{fig:figure7}{\bf Comparison between H plasma and RIE ribbon.} (a) Four-wire conductivity $\sigma$ as a function of gate voltage for two GNRs etched in the same encapsulated sample, fabricated as labeled. A series resistance is subtracted from each curve, consistent with the the number of squares between the ribbon and the contacts ($\sim200\,\Omega$ for blue and $\sim400\,\Omega$ for black curve). (b) AFM height image of H plasma etched GNR with a width of $\sim300\,$nm. Two Cr/Au edge contacts (not shown) are evaporated on each side of the ribbon after encapsulation and the black dashed regions are etched out to prevent short circuiting of the ribbon.}
\vspace{-1mm}
\end{figure}

\indent \textbf{Electronic Transport Through H Plasma Defined Constrictions.} In Figure\,\ref{fig:figure7}, we show transport measurements for narrower and shorter graphene constrictions fabricated in a different way. Prior to encapsulation, we define two round holes of small diameter into the graphene layer with Ar/O$_2$ plasma and expose these to the remote H plasma to create hexagons sandwiching a GNR with ZZ edges between them, see Figure\,\ref{fig:figure7}b. Here, the round seed hole is relatively small, only about $100$\,nm in size, and most of the $\sim 500$\,nm hexagon was etched with the H plasma process in an exposure of about 5 hours. For the Fabry-P\'{e}rot sample, in contrast, only a relatively small amount of H plasma etching was performed (about 1 hour), enlarging the Ar/O$_2$ plasma defined structures only slightly. It would be interesting to compare in transport experiments ribbons with long and short H plasma exposure, even tough no time dependence of the edge quality was observed on the Raman samples (see Figure\,\ref{fig:figure3}). It is plausible that longer exposure has a healing effect on the edge, such that it removes defects more efficiently and creates less disordered atomic arrangements. Plus, for longer exposures the ribbon direction is solely determined by the graphene lattice since the etching starts from a round defect and evolves naturally to a hexagon with edges along the ZZ crystal axis. In the Fabry-P\'{e}rot sample (short exposure), a small misalignment between the overall ribbon direction and the ZZ crystal axis may remain after the H plasma etching. Also, the e-beam defined circles are clearly visible as an elevated region. Such regions are known to appear after H plasma exposure and have been observed in many samples\cite{hug2017anisotropic}. However, further investigation is required to better understand these features.

Figure\,\ref{fig:figure7} shows an example of such a ZZ GNR, with 4-wire conductivity as a function of global back gate voltage plotted in panel a, comparing  a H plasma defined ribbon (blue) with a RIE defined ribbon (black). Both ribbons are fabricated on the same graphene flake, allowing direct comparison. As seen, the two curves are very similar, resulting in comparable mobilities, and no plateaus of quantized conductance are clearly evident. We note that for most of the gate voltage range the mean free path is larger than $500$\,nm in bulk, determined from a separate Hall bar sample, see SOM S6. Even tough the ribbons are about a factor of 2 shorter than the bulk mean free path, conductance quantization is not observed. Thus, we conclude that the edges are the dominant source of scattering, irrespective of whether they are defined with H plasma or RIE. We note that none of the wiggles seen in these conductivity traces obviously develop into a conductance plateau even under application of a magnetic field.

\subsection{Conclusion}

In conclusion, we have found that H plasma defined hexagons on graphite did not show any D-peak intensity, and thus seem to display high quality ZZ edges. In contrast, in SL graphene on hBN, a relatively large D-peak is seen on H plasma defined edges. Polarization dependent Raman measurements revealed an edge configuration consisting of approximately $60\,$\% ZZ and $40\,$\% AC-30$^\circ$ segments. Valley-isospin oscillations in quantum transport are again consistent with edge disorder with $40\,$\% AC-30$^\circ$ segments. Moreover, transport through narrow graphene constrictions showed similar mobilities for RIE defined edges and H plasma edges. Interestingly, bulk graphene shows high electronic quality after H plasma exposure, manifested in Fabry-P\'{e}rot resonances. Thus, exposure of graphene to the remote H plasma is an excellent cleaning method, since it removes residues very efficiently without degrading the quality of the graphene crystal lattice. This is further confirmed by Raman spectroscopy (no D-peak in the bulk), AFM (very clean surfaces without PMMA residues) and electronic transport measurements (high electronic mobility).

Further investigations are needed to identify possible origins of the disorder such as e.g. hBN-graphene interactions where the relative rotation angle could play a role\cite{xue2011scanning,wallbank2015moire}, or a too high thermal energy leading to edge reconstruction\cite{ThermalDynamics}. Thus, further optimization of the H plasma etching process is required in order to obtain high quality crystallographic graphene edges.

\subsection{Methods}

\textbf{AFM specifications:} Two different AFM instruments were used for the measurements presented in this work. The data shown in Figure\,\ref{fig:figure1}a to c was obtained by means of ambient qPlus based atomically resolving AFM\cite{wastl2013optimizing}, namely with a quartz force sensor with resonance frequency $f_0=33\,$kHz, stiffness $k=1800\,$N/m and quality factor $Q=3000$. Coarse topography images in Figure\,\ref{fig:figure1}a and b were acquired in the frequency-modulated mode while Figure\,\ref{fig:figure1}c shows an atomically resolved frequency shift image acquired in the constant height mode. The details of the setup are described elsewhere\cite{morita2015noncontact}. For all other AFM data a Bruker Dimension 3100 was used. All data measured with this instrument was acquired in intermittent contact mode (amplitude modulated).

\textbf{H plasma parameters:} The following parameters were used for the exposures of all samples presented in this work: $T=400\,$\degC, $p=1.7\,$mbar, H$_2$ gas flow of $20\,$SCCM. The details of the setup are described elsewhere\cite{hug2017anisotropic}. Anisotropic etching in a H plasma was also achieved using a $2.4\,$GHz solid state generator from MKS with a power of $100\,$W - $300\,$W and a flow of $100\,$SCCM; the pressure and temperature values were the same as stated above.

\textbf{Raman microscope:} The Raman measurements were acquired with a WITEC alpha300 Raman system. The wavelength of the He-Ne laser was $633\,$nm and the used objective was 100x with $NA=0.9$. The laser power was set to $1.5\,$mW or below for all measurements. This power is low enough to exclude any laser induced structural changes (see SOM S2). To extract the peak heights, we first subtract the background and fit single Lorentzains (for the G and 2D-peaks) and a double Lorentzian for the D and hBN peaks.

\textbf{Sample fabrication:} We used graphite flakes from NGS Naturgraphit GmbH. SL flakes were obtained by the scotch tape method\cite{novoselov2004electric} and transferred on top of hBN crystals by the wet transfer technique described in ref. \onlinecite{dean2010boron}. High quality hBN crystals\cite{taniguchi2007synthesis} were exfoliated on top of a p++ doped Si wafer covered by $300\,$nm SiO$_2$ following the same scotch tape method. For the sample in Figure\,\ref{fig:figure7}, we used an hBN flake from HQ-Graphene. After H plasma etching, the electronic transport samples were in addition encapsulated by a top hBN flake to ensure high cleanliness and stability of the devices. To cut the hBN/graphene/hBN stack in order to shape it or define control ribbons an SF$_6$/Ar/O$_2$ gas mixture was used in an RIE process. Cr/Au side-contacts were fabricated following ref. \onlinecite{wang2013one} with an additional RIE step with CHF$_3$/O$_2$ gas prior to metalization. The GNR presented in Figure\,\ref{fig:figure5} is $600\,$nm wide and measures $1.6\,\mathrm{\mu m}$ in length between the source and drain contacts. The bottom and top hBN layers have a thickness of $42\,$nm and approx. $35\,$nm, respectively.

\textbf{Electronic transport measurements:} Standard low frequency lock-in measurements were performed in a variable temperature insert at $T=4\,$K.

\textbf{Quantum transport simulations:} All quantum transport simulations are done with the following parameters. Magnetic field is fixed at $B=16\,$T, the distance between two \textit{pn}-interfaces is set to $20\,$nm, the ribbon is about $40\,$nm wide, and the smoothness of the \textit{pn} and \textit{np} regions is approximately $5\,$nm. Device scaling is necessary due to computational capacity reasons and the strong magnetic field is required to ensure that the Landau levels are well developed in such a narrow ribbon. All calculations with disorder are averaged over 200 different configurations of Anderson-type disorder potential. All calculations are done at zero temperature.



%

\subsection{Acknowledgments}

We would like to thank J. Roche, C. Handschin and J. Overbeck for technical advice and helpful discussions. We acknowledge support from the Swiss Nanoscience Institute (SNI), NCCR QSIT, Swiss NSF, Taiwan Minister of Science and Technology (MOST), Grant No. 107-2112-M-006 -004 -MY3 and financial support from Innosuisse. We also thank Korbinian Pürckhauer, Anja Merkel, Florian Griesbeck and the machine shop of U Regensburg for building, assembling and testing the ambient qPlus based atomically resolving AFM and the Deutsche Forschungsgemeinschaft for funding under GRK 1570. Growth of hexagonal boron nitride crystals was supported by the Elemental Strategy Initiative conducted by the MEXT, Japan and the CREST (JPMJCR15F3), JST.

\bibliographystyle{myStyle_mr}

\end{document}